\documentclass[sigconf]{acmart}
\AtBeginDocument{%
  }

\setcopyright{acmlicensed}

\copyrightyear{2026}
\acmYear{2026}
\setcopyright{cc}
\setcctype{by}
\acmConference[KDD 2026] {Proceedings of the 32nd ACM SIGKDD Conference on Knowledge Discovery and Data Mining V.1}{August 9--13, 2026}{Jeju Island, Republic of Korea.}
\acmBooktitle{Proceedings of the 32nd ACM SIGKDD Conference on Knowledge Discovery and Data Mining V.1 (KDD 2026), August 9--13, 2026, Jeju Island, Republic of Korea}
\acmISBN{979-8-4007-2258-5/2026/08}
\acmDOI{10.1145/3770854.3780171}

\usepackage{algorithm}           
\usepackage{algpseudocode}
\usepackage{makecell}
\usepackage{hyperref}       
\usepackage{url}            
\usepackage{booktabs}       
\usepackage{amsfonts}       
\usepackage{nicefrac}       
\usepackage{microtype}      
\usepackage{xcolor}         
\usepackage{multirow}  
\usepackage{graphicx}
\usepackage{amsmath}
\usepackage{enumitem,xfrac} 
\usepackage{multirow}
\usepackage{float} 
\usepackage[table]{xcolor}   
\usepackage{amsmath}

\newcommand*{\ourmodel}{MixRAG}
\newcommand*{\ourdataset}{DocRAGLib}
\newcommand*{\taskname}{Heterogeneous Document RAG}

\newcommand*{\CorpusModule}{Heterogeneous Document Representation Module}
\newcommand*{\RetrievalModule}{Cross-Modal Retrieval Module}
\newcommand*{\RecapModule}{Multi-Step Reasoning Module}

\newcommand{\cqy}[1]{\textcolor{black}{#1}}

\newcommand{\zc}[1]{\textcolor{black}{#1}}

\begin{document}


\title{Mixture-of-RAG: Integrating Text and Tables with Large Language Models}

\author{Chi Zhang}
\authornote{Chi Zhang and Qiyang Chen contribute equally to this work.
Beijing Institute of Technology is the primary corresponding institution for this work.
The main work presented in this paper was completed in July 2022.}
\email{chi_zhang@bit.edu.cn}
\orcid{0009-0000-0323-0715}
\affiliation{
  \institution{Beijing Institute of Technology}
  \city{Beijing}
  \country{China}
}

\author{Qiyang Chen}
\email{qiyangchen@bit.edu.cn}
\orcid{0009-0007-0913-1970}
\authornotemark[1]
\affiliation{
  \institution{Beijing Institute of Technology}
  \city{Beijing}
  \country{China}
}

\author{Mengqi Zhang}
\authornote{Corresponding Author.}
\email{mioplk701@gmail.com}
\orcid{}
\affiliation{
  \institution{Zhejiang Gongshang University}
  \city{Hangzhou}
  \state{Zhejiang}
  \country{China}
}



\renewcommand{\shortauthors}{Chi Zhang, Qiyang Chen, \& Mengqi Zhang}

\begin{abstract}
Large language models (LLMs) achieve optimal utility when their responses are grounded in external knowledge sources.
However, real-world documents, such as annual reports, scientific papers, and clinical guidelines, frequently combine extensive narrative content with complex, hierarchically structured tables. 
While existing retrieval‑augmented generation (RAG) systems effectively integrate LLMs’ generative capabilities with external retrieval-based information, their performance significantly deteriorates especially processing such heterogeneous text-table hierarchies. 
To address this limitation, we formalize the task of \taskname{}, which requires joint retrieval and reasoning across textual and hierarchical tabular data. 
We propose \ourmodel{}, a novel three‑stage framework: (i) hierarchy row-and-column-level (H-RCL) representation that preserves hierarchical structure and heterogeneous relationship, (ii) an ensemble retriever with LLM-based reranking for evidence alignment, and (iii) multi‑step reasoning decomposition via a RECAP prompt strategy. 
To bridge the gap in available data for this domain, we release the dataset \ourdataset{}, a 2k‑document corpus paired with automatically aligned text‑table summaries and gold document annotations. 
The comprehensive experiment results demonstrate that \ourmodel{} boosts top‑1 retrieval by 46\%  over strong text‑only, table‑only, and naive‑mixture baselines, establishing new state-of-the-art performance for mixed-modality document grounding. 
\end{abstract}




\maketitle

\section{Introduction}
\label{Introduction}

With the rapid development of large language models (LLMs)~\cite{llama2,mistral7b}, the reliability of their response substantially depends on the extent to which they can be grounded in verifiable evidence. 
Retrieval-Augmented Generation (RAG) has emerged as an effective strategy that combines the generative power of LLMs with external retrieval-based information. 
Recent research has extended the application of RAG to various scenarios, including knowledge graph~\cite{Graph_MDQA, matsumoto2024kragen}, database~\cite{biswal2024tag}, and multi-modal data~\cite{multi-moda_PreFLMR,Multi-moda-sigir1}. When the external source is plain text or simple tables, RAG systems retain precision and fluency. However, real-world data, including financial statements, clinical guidelines, research articles, and governmental reports, all interleave complex hierarchical tables whose rows and columns nest multiple semantic layers. The crucial metrics numbers, such as earnings per share, dosage limits, and experimental results, are embedded within these tables, and the relationships among those numbers are defined by the headers that span multiple rows or columns. Therefore, the research on the \taskname{} remains nascent. The \taskname{} aims to efficiently retrieve and synthesize relevant information from large, complex documents by integrating textual and hierarchical tabular data with the power of LLMs for more comprehensive generation.

\begin{figure}
    \centering
    \includegraphics[width=0.9\linewidth]{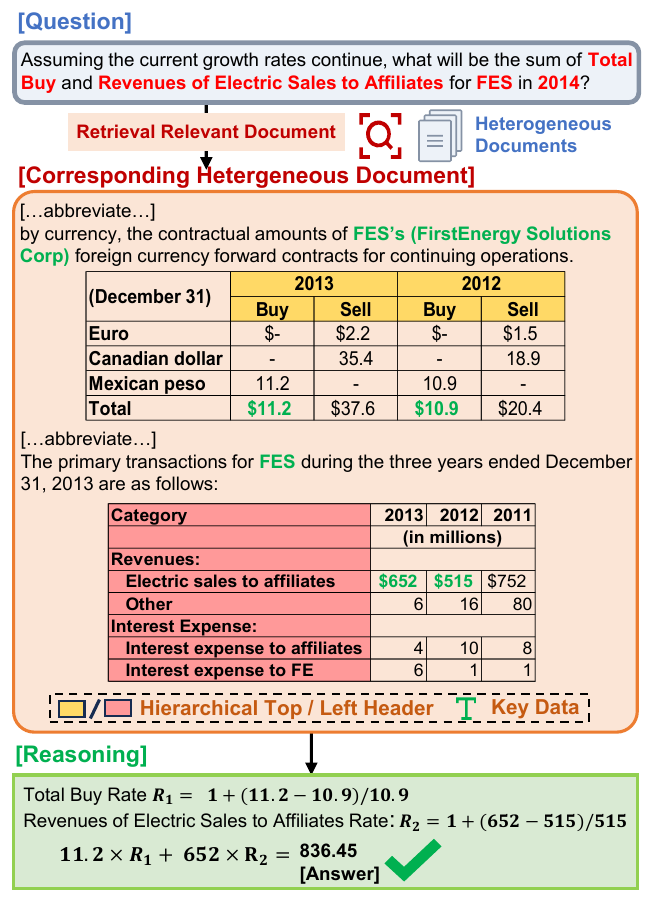}
    \caption{An Example of \taskname{} Task}
    \label{intro_figure}
\end{figure}
The \taskname{} task involves retrieving the most relevant documents from extensive document repositories based on a specific question and generating the corresponding answer. 
However, it is challenging due to the complexity and scale of real-world repositories, which often contain numerous indistinguishable documents with intricate structures.
For example, Statistics Canada offers thousands of statistical reports across subjects, including 1,761 on ``Labour'' and 1,132 on Population and ``demography'', with semantically similar content often combining textual descriptions and hierarchical tables.

Some novel RAG techniques ~\cite{izacard2020distilling,borgeaud2022improving,wang-etal-2024-dapr,islam-etal-2024-open} have been explored to improve retrieval accuracy by enhancing RAG frameworks. While effective in text-based tasks, these methods often fail to capture the critical table information within retrieved documents.
Other table-focused RAG techniques~\cite{biswal2024tag,roychowdhury2024eratta,cheng2024retrievalonmultimodal,matsumoto2024kragen} lack seamless integration of table data with text, resulting in unaligned representations and difficulty handling complex hierarchical structures.
As shown in Figure~\ref{intro_figure}, when retrieval relies on the hierarchical structure of a table, such methods often break down the complex structure, leading to information loss, i.e., disruption of critical dependencies (e.g., 2012-2014 trends), and a reduced ability to capture critical relationships within the data. 
Additionally, while well-established document-based QA methods~\cite{ProTrix,srivastava-etal-2024-evaluating,TACR} have demonstrated relatively high accuracy, they still struggle with complex tasks involving documents with both long text and hierarchical tables~\cite{zhao-etal-2024-docmath}. 
Overall, although existing methods have provided preliminary solutions for \taskname{}, achieving more precise information retrieval and integration still faces numerous challenges: \textbf{(i) Data Scarcity.} The publicly available benchmarks lack a cleaned, integrated corpus pairing narrative text with hierarchical tables.
\textbf{(ii) Structured Representation.} The unified representation must preserve row‑column hierarchies while remaining consistent with text. As shown in Figure~\ref{intro_figure}, the key terms ``\textit{Total Buy}'' and ``\textit{Revenues of Electric Sales}'' are strongly linked to the hierarchical structure within the heterogeneous document, and the information contained within tables is dense, highlighting the challenge of representing structured data in such documents. 
\textbf{(iii) Retrieval Precision.} 
The traditional semantic similarity-based retrieval methods often return text chunks that, while semantically related to the question, may not provide valid evidence for reasoning. For example, as illustrated in Figure~\ref{intro_figure}, existing methods may retrieve documents containing the exact year `2014' from the query, while missing the relevant historical data from 2012 and 2013 that the query actually seeks.
\textbf{(iv) Reasoning Accuracy.} Effective retrieval requires hierarchical alignment between unstructured text and structured tabular data, necessitating the ability to process and reason across heterogeneous documents accurately (e.g., deriving 2012-2013 context from a 2014 query and multi-step computation). 

To address the above challenges in \taskname{}, we propose the \textbf{\ourdataset{}} dataset and \textbf{\ourmodel{}}\footnote{Our dataset, code, and prompts are available in GitHub at \url{https://github.com/ChiZhang-bit/Mixture-of-RAG}.} framework.
Specifically, \ourdataset{} consists of a document repository with 2,178 documents, each of which combines text and hierarchical tables, as well as 4,468 QA pairs. \ourdataset{} is collected from reliable sources and cleaned to ensure consistency across document types and formats. To the best of our knowledge, \ourdataset{} is the first dataset that bridges the gap of heterogeneous documents. \ourmodel{} framework consists of three modules: \CorpusModule{}, \RetrievalModule{}, and \RecapModule{}.
Specifically, the \CorpusModule{} employs a hierarchical row-and-column-level (H-RCL) table summarization method to capture the structure and content of complex tables. It generates representations that preserve table structures and are optimized for retrieval, addressing the challenge of effectively representing table information in heterogeneous documents.
The \RetrievalModule{} overcomes the limitations of relying solely on semantic similarity retrieval by adopting a two-stage approach. First, the ensemble retrieval stage combines BM25 with the semantic understanding of embedding retrieval to filter candidate documents. Second, the LLM-based reranking stage utilizes the contextual reasoning ability of LLMs to identify the most relevant document, improving the accuracy and comprehensiveness of retrieval.
The \RecapModule{} introduces the RECAP prompt strategy, which decomposes complex reasoning into sub-tasks and leverages external calculators to manage mathematical operations during the reasoning process. \ourmodel{} significantly improves answer accuracy and effectively addresses the challenges of multi-step reasoning and complex calculations in heterogeneous documents.

In summary, our main contributions are as follows:
\begin{itemize}
    \item We propose a new dataset, \ourdataset{}, which provides a high-quality benchmark for heterogeneous information retrieval tasks.
    \item We design the \ourmodel{} framework, which consists of three components: the \CorpusModule{}, the \RetrievalModule{}, and the \RecapModule{}. 
    \item We conduct extensive experiments to validate the superior performance of the \ourmodel{} framework over baseline methods, with ablations verifying the effectiveness of each module and scalability analysis showing the framework's robustness across different corpus sizes.
\end{itemize}

\section{Related Work}
In this section, we present an overview of relevant research from two perspectives: RAG and Document QA.
\label{related_work}
\subsection{RAG}
RAG has significantly enhanced language models, particularly in knowledge-intensive tasks involving document data. This section presents existing work from two main aspects: advancements in RAG techniques~\cite{borgeaud2022improving, wang-etal-2023-query2doc, wang-etal-2024-dapr, izacard2020distilling, yan2024crag, islam-etal-2024-open, asai2023selfrag, cheng-etal-2024-call} and its applications across different modalities~\cite{biswal2024tag, cheng2024retrievalonmultimodal, matsumoto2024kragen, roychowdhury2024eratta, hu2024mrag}.

Advancements in RAG techniques aim to improve retrieval accuracy, efficiency, and robustness in knowledge-intensive tasks. 
Izacard et al. \cite{izacard2020distilling} use knowledge distillation to train retrievers without annotated query-document pairs, relying on reader model attention for synthetic labels, though it depends on the reader model's quality. RETRO~\cite{borgeaud2022improving} combines document chunks retrieved by local similarity with preceding tokens in auto-regressive models but faces challenges with long documents. DAPR \cite{wang-etal-2024-dapr} addresses this by integrating hybrid retrieval with BM25 and contextualized passage representations for long document passages. Open-RAG~\cite{islam-etal-2024-open} improves multi-hop reasoning and retrieval accuracy by transforming dense LLMs into the parameter-efficient sparse mixture of expert models. However, these methods primarily focus on text-based RAG, with limited progress in handling mixed modalities, which pose unique retrieval and reasoning challenges.

Recent advancements in RAG have extended to multimodal data beyond text. TAG~\cite{biswal2024tag} and ERATTA~\cite{roychowdhury2024eratta} introduce RAG frameworks for table-structured data. Chen et al.~\cite{chen-etal-2024-table} propose a method combining join relationship discovery and mixed-integer programming-based re-ranking for table retrieval. RAGTrans~\cite{cheng2024retrievalonmultimodal} integrates textual and multimedia information for enhanced representation and retrieval. KRAGEN~\cite{matsumoto2024kragen} introduces a knowledge graph-enhanced RAG framework that uses graph-based retrieval to improve factual consistency and reasoning in biomedical problems. MRAG-Bench \cite{hu2024mrag} focuses on vision-centric RAG, highlighting cases where visual knowledge outperforms textual information, especially for vision-language models. However, none of these approaches address RAG for the combined modality of tables and text within documents, a gap that \ourmodel{} specifically addresses.

\subsection{Document QA}
Document QA refers to the task of extracting or generating answers based on the content of a given document. This section reviews existing approaches to Document QA from two primary perspectives: methods leveraging pre-trained language models~\cite{chen-etal-2020-hybridqa,zhu-etal-2021-tat,cheng-etal-2022-hitab,zhao-etal-2022-multihiertt,10.1007/978-3-031-44693-1_46} and methods utilizing LLM~\cite{luo2023hrot,srivastava-etal-2024-evaluating,CoT,PoT,zhao-etal-2024-docmath}.

In Document QA, pre-trained NLP models are fine-tuned to process document-question pairs, enabling them to extract information or generate answers based on the content. HybridQA~\cite{chen-etal-2020-hybridqa} and TAT-QA~\cite{zhu-etal-2021-tat} apply QA tasks to text-table hybrid documents, but their tables are typically flat. HiTab~\cite{cheng-etal-2022-hitab} focuses on hierarchical table-based QA tasks but struggles with hybrid data combining text and hierarchical tables. Zhao et al.~\cite{zhao-etal-2022-multihiertt} introduce the MultiHiertt benchmark, containing both text and hierarchical tables and propose MT2Net for this task. MT2Net extracts facts from hybrid documents and performs reasoning, but it has high content requirements and depends heavily on table cell descriptions for accurate reasoning. NAPG~\cite{10.1007/978-3-031-44693-1_46} outperforms MT2Net on MultiHiertt, yet existing NLP models still face challenges with multi-step reasoning, especially in complex mathematical computations.

With the emergence of advanced LLMs, several studies apply them to question-answering tasks with hybrid documents. Luo et al. \cite{luo2023hrot} introduce the HRoT strategy for text-table hybrid QA, achieving better results than existing pre-trained NLP models on the MultiHiertt dataset. Srivastava et al. \cite{srivastava-etal-2024-evaluating} propose EEDP, designed for semi-structured documents, and compare it with other prompting methods (such as CoT~\cite{CoT} and PoT~\cite{PoT}) across various QA datasets. Zhao et al.~\cite{zhao-etal-2024-docmath} evaluate the numerical reasoning abilities of 27 LLMs using CoT and PoT on text-table hybrid documents. Despite the progress of advanced prompting strategies in guiding LLMs for hybrid document QA, there remains room for improvement compared to human performance, highlighting the need for more efficient prompting techniques.

\section{Preliminary}
\label{Preliminary}

In this section, we formally define the \textbf{\taskname{}} task and detail the collection, processing, and cleaning procedures for our dataset, \ourdataset{}.

\subsection{Task Definition}
\label{sec: formolize}
In this section, we formally define key concepts and our \textbf{\taskname} task.

1) \underline{\textit{Document Corpus.}}
A comprehensive collection of documents $ \mathbf{C} = \{ D_1, D_2, \dots, D_{|\mathbf{C}|} \} $ where each document $ D_i \in \mathbf{C}  $ consists of multiple text segments $ \mathbf{P} = \{ P_1, P_2, \dots, P_{|\mathbf{P}|} \} $ and several complex hierarchical tables $\mathbf{T} = \{ T_1, T_2, \dots, T_{|\mathbf{T}|} \}$, referred to as \textit{heterogeneous document}. Each hierarchical table is defined as a tuple $T=\{{H}_t, {H}_l, d\}$, where ${H}_t$ denotes the top headers in hierarchical structure of table $T$, ${H}_l$ denotes the left headers in hierarchical structure of table $T$ (example of hierarchical headers shown in Figure~\ref{intro_figure}), and $d$ denotes the data cells of table $T$. 

2) \underline{\textit{Question}.} A question $Q$ posed by the user, for which there exists a unique document of highest relevance $D^* \in \mathbf{C} $ in the corpus.

Given a question $Q$ and a document corpus $\mathbf{C}$, our task is to retrieve the most relevant document $D^*$ and generate a comprehensive answer $A$ based on the information contained within $D^*$. Formally, the \textbf{\taskname{}} task can be described by the following function:
\begin{equation}
\small
A = \mathcal{F}(Q, \mathbf{C}) = \text{Inference}\left( Q, D^* \right), \text{where } D^*=\text{Retrieve}(Q, \mathbf{C})
\end{equation}
where $\text{Retrieve}(Q, \mathbf{C})$ is the function that retrieves the most relevant document $D^* \in \mathbf{C}$ based on the question $Q$, and $\text{Inference}(Q, D^*)$ is the generation function that returns an accurate answer $A$ by leveraging the information contained within $D^*$.   

\subsection{Dataset}

Existing datasets are predominantly tailored for single-document RAG tasks, limiting their applicability to scenarios requiring the processing of complex document corpora. To bridge this gap, we introduce \textbf{\ourdataset{}}, a comprehensive dataset specifically tailored for multi-document RAG tasks.

\begin{figure}
    \centering
    \includegraphics[width=1\linewidth]{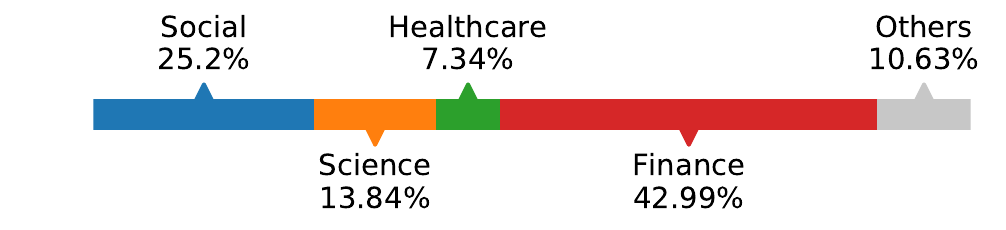}
    \caption{Distribution of domains in \ourdataset{}}
    \label{fig:statistic}
\end{figure}

The documents of \ourdataset{} are derived from two primary sources: existing public single-document QA datasets and web-scraped heterogeneous data. 
First, we incorporate samples from MultiHiertt~\cite{zhao-etal-2022-multihiertt}, where each document comprises textual content and hierarchical tables (averaging 3.89 tables per document). 
Second, we scrape contextual information utilizing webpage links provided in HiTab~\cite{cheng-etal-2022-hitab}, reconstructing heterogeneous documents that contain both text and tables. 
This hybrid collection strategy ensures broad domain diversity, covering multiple fields such as finance, social, and healthcare, as shown in Figure~\ref{fig:statistic}.

In total, \ourdataset{} contains 2,178 documents, combining heterogeneous text and table content, and includes 4,468 QA pairs meticulously designed based on these documents. Further details regarding Data Cleaning and additional statistical analysis are elaborated in Appendix~\ref{Dataset_Appendix}.


\section{Methodology}
\label{Methodology}
To address this problem, we introduce \textbf{\ourmodel{}}, a framework designed for the \taskname{} Task.
\begin{figure*}
    \centering
    \includegraphics[width=1\linewidth]{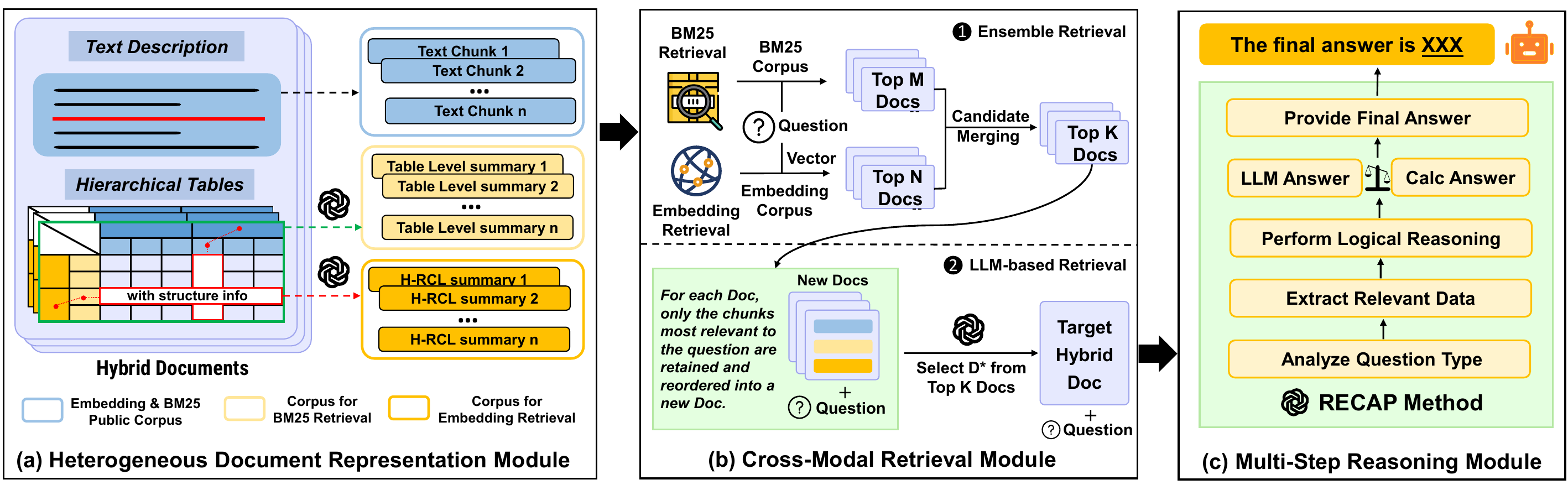}
    
    \caption{The Overview of our \ourmodel{} Framework}
    \label{fig:framework}
\end{figure*}
As depicted in Figure~\ref{fig:framework}, \ourmodel{} comprises three modules: \CorpusModule{}, \RetrievalModule{}, and \RecapModule{}.
The following sections show a detailed introduction to the three modules.

\subsection{\CorpusModule{}}
The \CorpusModule{} focuses on generating effective representations of heterogeneous documents, capturing both lengthy text and complex hierarchical tables to optimize retrieval performance.
A traditional approach, the Table Level Summary, involves providing the table's schema to the LLM and summarizing it. However, it oversimplifies the table, leading to the loss of key information and ignoring important structural details.
For effective representation, the \CorpusModule{} first introduces a row-and-column-level (RCL) table summary to represent tables. Then, leveraging the hierarchical structure of the tables, we extend this RCL table summary to capture the nested relationships and dependencies within the table data, ensuring a more comprehensive representation.

To better understand the proposed RCL table summary, we introduce some basic concepts related to table structure and paths.

For a hierarchical table, as we defined before, table $ T = \{ H_t, H_l, d \} $, where $ H_t $ represents the top headers, $ H_l $ represents the left headers, and $ d $ represents the data cells. Specifically, we can define:
{\small
\begin{equation}
    \left\{
    \begin{aligned}
        H_t &= \{ H_t^r \mid r = 1, \dots, R_t \} \\
        H_l &= \{ H_l^r \mid r = 1, \dots, R_l \}
    \end{aligned}
    \right.
\end{equation}
}
where $ R_t $ denotes the number of hierarchical levels in the top headers, and $ R_l $ denotes the number of hierarchical levels in the left headers.

For each hierarchical level in the top and left headers, we define:
{\small
\begin{equation}
    \left\{
    \begin{aligned}
        H_t^r &= \{ h_t^r(j) \mid j \text{ is the index of the header at top-$r$ level} \} \\
        H_l^r &= \{ h_l^r(i) \mid i \text{ is the index of the header at left-$r$ level} \}
    \end{aligned}
    \right.
    \label{header}
\end{equation}
}
where $ h_t^r(j) $ represents the header at level $ r $ in the top header hierarchy, while $ h_l^r(i) $ represents the header at level $ r $ in the left header hierarchy. As shown in Figure~\ref{fig:corpus}, these headers form multi-level paths used to locate specific data cells within the table. To pinpoint a particular data cell $ d_{ij} $, we define the paths from both the left and top headers, $ P_l(i) $ and $ P_t(j) $, respectively, which together uniquely identify the location of the data cell:
{\small
\begin{equation}
    \left\{
    \begin{aligned}
        P_l(i) &= h_l^1(i_1) \to h_l^2(i_2) \to \dots \to h_l^{R_l}(i_{R_l}) \\
        P_t(j) &= h_t^1(j_1) \to h_t^2(j_2) \to \dots \to h_t^{R_t}(j_{R_t})
    \end{aligned}
    \right.
    \label{path}
\end{equation}
}
These paths capture the hierarchical relationships between the cells in the table, allowing us to pinpoint any data cell $ d_{ij} $ by referencing the corresponding paths in the left and top headers.

Below, we provide a detailed formal definition of the proposed RCL table summary, outlining its structure and how it incorporates hierarchical relationships for enhanced table representation.

\subsubsection{General RCL Table Summary}
In the case of a general table, where multi-level hierarchies and complex relationships within the table are not considered, we summarize each row and column independently. By disregarding the hierarchical structure, the table is reduced to a flat, 1-level representation, where both the top and left headers are flattened.

Formally, when the hierarchical structure is ignored, $H_t$ and $H_l$ are reduced to:
{\small
\begin{equation}
    \left\{
    \begin{aligned}
        H_t &= \{ h_t(1), h_t(2), \dots, h_t(n) \mid n \text{ is \# columns} \} \\
        H_l &= \{ h_l(1), h_l(2), \dots, h_l(m) \mid m  \text{ is \# rows} \}
    \end{aligned}
    \right.
\end{equation}
}
Similarly, the paths simplify to $P_l(i) = h_l(i)$ and $P_t(j) = h_t(j)$. In this generalized case, each row and column is represented by a left and top header, respectively. To summarize the table at the general row and column level, we flatten the table's hierarchical structure and treat it as a single level. Specifically, for each row and column, the general RCL table summary can be formulated as:
{\small
\begin{equation}
\small
    \left\{
    \begin{aligned}
        S_{\text{row}} \{ h_l(i) \} &= \mathcal{G}(h_l(i), \bigcup_{j=1}^nh_t(j)) \\
        S_{\text{col}} \{ h_t(j) \} &= \mathcal{G}(h_t(j), \bigcup_{i=1}^mh_l(i))
    \end{aligned}
    \right.
\end{equation}
}
where $\mathcal{G}$ denotes the operation that generates the general RCL summary for both rows and columns. This operation condenses the information in each row and column, simplifying the table's hierarchical structure for retrieval.

\subsubsection{H-RCL Table Summary}


Hierarchical tables contain multiple levels of headers that form a complex hierarchy, enriching the data structure. In this context, we extend the General RCL table summary to handle multi-level structures, which capture the intricate dependencies between the table’s headers and data more effectively.

\begin{figure}
    \centering
    \includegraphics[width=1\linewidth]{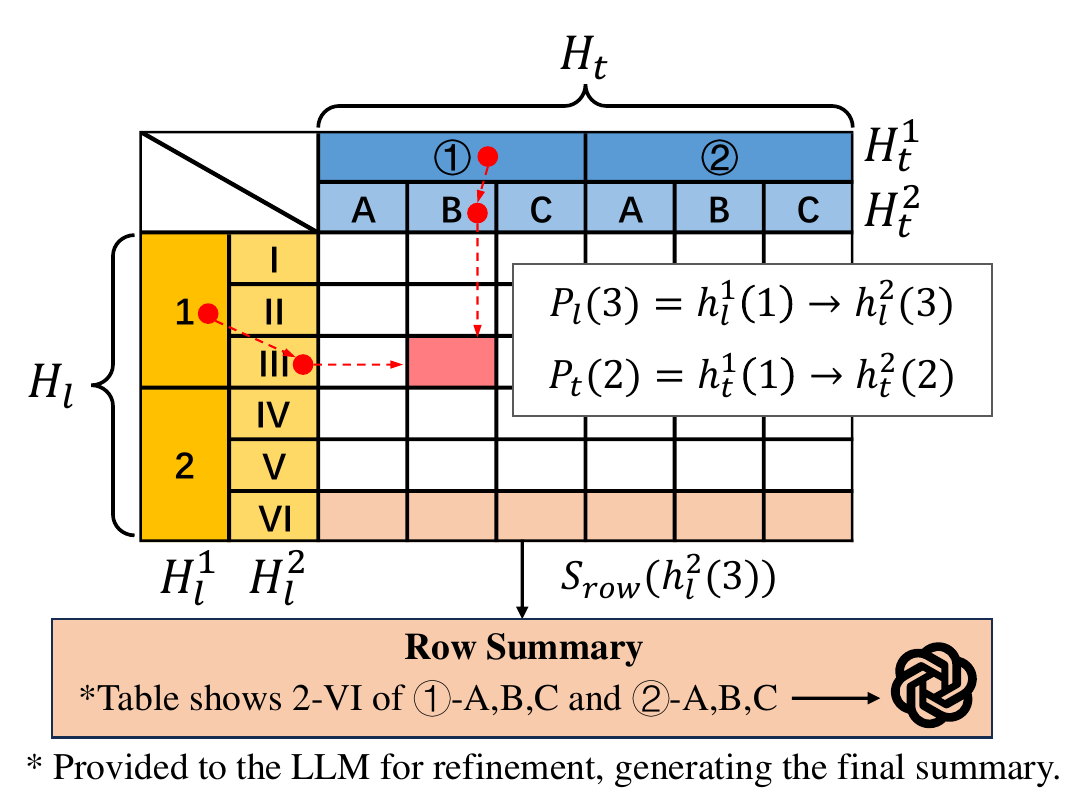}
    \caption{The Path and Hierarchical Levels in the Table}
    \label{fig:corpus}
\end{figure}

We propose the H-RCL table summary that preserves the nested relationships between rows and columns by leveraging the hierarchical paths of both top and left headers, as shown in Eq.~\ref{header} and Eq.~\ref{path}. In the following, we provide a formal definition of the hierarchical row and column summaries:
{\small
\begin{equation}
\small
    \left\{
    \begin{aligned}
        S_{\text{row}}(h_l^r(i)) &= \mathcal{G}(P_l^r(i), \bigcup_{j=1}^n P_t^{R_t}(j)) \\
        S_{\text{col}}(h_t^r(j)) &= \mathcal{G}(P_t^r(j), \bigcup_{i=1}^m P_l^{R_l}(i))
    \end{aligned}
    \right.
\end{equation}
}
The row-level summaries $ S_{\text{row}} $ capture the dependencies within the left headers, while the column-level summaries $ S_{\text{col}} $ reflect the relationships within the top headers. 
As shown in Figure~\ref{fig:corpus}, the $S_{\text{row}}$ aggregates information from both the left and top headers at their respective hierarchical levels. After aggregation, the information is provided to the LLM for refinement to generate the final summary while preserving the complex relationships within the table’s structure. These summaries jointly represent the hierarchical structure of the table. The overall table summary is defined as:
{\small
\begin{equation}
    S_{\text{table}} = \bigcup_{i,r} S_{\text{row}}(h_l^{r}(i)) \cup \bigcup_{j,r} S_{\text{col}}(h_t^{r}(j))
\end{equation}
}
By combining both row and column summaries, the table summary $ S_{\text{table}} $ offers a comprehensive representation of the hierarchical table, effectively maintaining the nested structure of the data.


\subsubsection{Passage Processing}

In addition to processing tables, we also address the handling of text passages within the heterogeneous document.
In this module, we focus on processing the passage at the sentence level.
\zc{To ensure that each sentence carries complete semantic meaning, we first apply coreference resolution using \textit{spaCy}, replacing referential expressions with their explicit noun phrase antecedents.
Then the document text is segmented into individual sentences using \textit{spaCy's} sentence tokenizer.
To reduce noise, we filter out overly short sentences (less than 5 words), which often lack meaningful content.
The resulting collection of meaningful sentences forms the textual portion of document corpus, supporting retrieval and downstream reasoning tasks.
}



\subsection{Cross-Modal Retrieval Module}
The Cross-Modal Retrieval Module is designed to identify the most relevant document $D^*$ from the document corpus for a given question. It consists of two key stages: Ensemble retrieval and LLM-based retrieval.
The ensemble retrieval stage combines BM25 and embedding-based retrieval to balance keyword matching with semantic understanding. BM25 captures exact term relevance, while embedding-based retrieval ensures deeper semantic alignment, enabling the retrieval of a comprehensive set of top-$k$ candidate documents. Specifically, we utilize two different types of corpora for the ensemble retrieval stage, tailored to align with the underlying mechanisms of embedding-based and BM25-based approaches. The ensemble retrieval stage conducts a coarse-grained
filtering of documents across the entire document corpus, ensuring that the
candidate documents provided to the LLM remain within its maximum input length constraints.
The LLM-based retrieval performs fine-grained logical reasoning tailored to the specific question and selected candidate documents, leveraging the contextual reasoning capabilities of the LLM to prioritize documents most relevant to the question. This approach ensures high precision in the final output.

\textbf{The Ensemble Retrieval Stage.} The ensemble retrieval stage is designed to generate a robust set of top-$k$ candidate documents by combining the complementary strengths of BM25 and embedding-based retrieval methods. Below, we detail the three key components of this stage: BM25 retrieval, embedding-based retrieval, and candidate merging. (1) \textbf{BM25 Retrieval.} We use BM25 to rank chunks from the BM25 corpus based on their relevance to a given question $Q$. Each chunk corresponds to a specific section of its document. BM25 emphasizes exact keyword matches and considers term frequency within a chunk while normalizing for document length to ensure fair comparisons. By prioritizing terms that are frequent in the chunk but rare across the corpus, BM25 effectively retrieves the top-$n$ candidate documents that are most aligned with the question.
As the RCL method may over-aggregate keywords and BM25's common-term bias, we use the corpus constructed with the table summary method for retrieval at this stage. 
(2) \textbf{Embedding-based Retrieval.} Simultaneously, embedding-based retrieval identifies the top-$m$ candidate documents from the embedding corpus by computing semantic similarity between the question $Q$ and each chunk. Both the question and chunks are represented as dense vectors generated by the embedding model. The embedding-based retrieval method captures semantic nuances and retrieves chunks that may not share explicit terms with the question but are contextually relevant.
In this retrieval stage, we utilize the corpus constructed using the H-RCL method, ensuring that the embedding model can more precisely capture information semantically similar to the question within hierarchical tables. (3) \textbf{Candidate Merging.} After retrieving the top-$n$ candidates from BM25 and top-$m$ from embedding-based retrieval, the two sets are merged into a final top-$k$ set with duplicates removed. This approach combines the strengths of both methods, ensuring a diverse and semantically rich candidate pool.

The ensemble approach combines the strengths of BM25 and embedding-based retrieval to create a diverse set of candidate documents. BM25 ensures accurate keyword-based matches, while embedding-based retrieval adds semantic depth and flexibility. By merging, the ensemble retrieval stage provides high recall and relevance, laying a strong foundation for the subsequent LLM-based retrieval stage.

\textbf{The LLM-based Retrieval Stage.} After the preceding Ensemble Retrieval Stage, the Cross-Modal Retrieval Module extracts the top-$ k $ candidate documents from the original document corpus that are most similar to the question. To further filter out the target document that is most relevant to the question, we introduce LLM as an expert in similarity analysis. Specifically, each question alongside its corresponding candidate documents is presented to the LLM, which is tasked with determining the single document most relevant to the question. However, since each document is typically lengthy, simultaneously inputting the top-$ k $ documents into the LLM often exceeds the model's maximum input token limit.
Moreover, the target document may contain a large amount of irrelevant information, which can reduce the LLM's accuracy when processing the entire content. To mitigate these challenges, we design the input space by setting a variable filtering threshold, $\theta$, and selecting the top-$\theta$ information chunks most similar to the question from all documents. 
In this process, an external function ensures that $\theta$ is set to its maximum value, provided that the token length of these chunks does not exceed the model's input limit.
These selected chunks are then consolidated into a single collection:
{\small
\begin{equation}
S = \text{SelectTop}(Q, D, \theta)
\end{equation}
}
where $ Q $ represents the original question. $ D $ denotes the set of all documents. $ \theta $ is the filtering threshold. $ S $ represents the set of selected information chunks.

Subsequently, for each document $ D_i $ in the top-$ k $ documents, only the information belonging to this collection is retained, and the filtered information is recombined into a ``new document'' $ D'_i $ to replace the original document:
{\small
\begin{equation}
D'_i = \text{Recombine}(S, D_i)
\end{equation}
}
Finally, we input the question content $ Q $ and its corresponding top-$ k $ new documents $ \{D'_1, D'_2, \ldots, D'_k\} $ into the LLM for similarity analysis, thereby selecting the most relevant document $ D^* $ among the top-$ k $ documents.



\subsection{\RecapModule{}}
\begin{table}[htbp]
\small
    \centering
    \caption{\zc{Performance Comparison with Various Baselines}}
    \begin{tabular}{l|c|c|c|c|c}
    \toprule
     \textbf{Method} & \multicolumn{4}{c|}{\textbf{HiT@K(\%)}} & \multirow{2}{*}{\textbf{EM*(\%)}} \\
    \cmidrule(lr){2-5}
     \textbf{\quad (\textit{- LLM})} & \textbf{K=1} & \textbf{K=3} & \textbf{K=5} & \textbf{K=10} & \\
    \midrule
    Standard RAG & 1.59 & 3.39 & 5.38 & 12.55 & 0.72|0.92|0.51$\spadesuit$ \\
    \midrule
    Table Retrieval & 37.05 & 52.99 & 59.96 & 64.94 & 22.44|25.41|19.06$\spadesuit$ \\
    \midrule
    \multicolumn{6}{l}{Langchain} \\
    \textit{- GPT-4o-mini} & 20.70 & 35.55 & 46.72 & 49.39 & 7.89 \\
    \textit{- GPT-4o} & 23.90 & 41.04 & 48.21 & 52.19 & 13.01 \\
    \textit{- Mistral-Nemo} & 17.93 & 31.15 & 43.55 & 48.05 & 7.07 \\
    \midrule
    \multicolumn{6}{l}{Self-RAG} \\
    \textit{- GPT-4o-mini} & 26.33 & 41.29 & 48.67 & 56.45 & 20.70 \\
    \textit{- GPT-4o} & 28.29 & 45.02 & 50.00 & 55.98 & 21.41 \\
    \textit{- Mistral-Nemo} & 19.88 & 37.19 & 40.68 & 54.30 & 15.58 \\
    \midrule
    \multicolumn{6}{l}{RAG-Critic} \\
    \textit{- GPT-4o-mini}  & 25.41 & 42.93 & 47.75 & 59.63 & 22.23 \\
    \textit{- GPT-4o}       & 29.00 & 46.11 & 50.51 & 60.55 & 25.31 \\
    \textit{- Mistral-Nemo} & 21.31 & 37.40 & 40.88 & 55.74 & 19.36 \\
    \midrule
    \multicolumn{6}{l}{\textbf{\ourmodel{}}} \\
    \textit{- GPT-4o-mini} & 48.46 & 69.36 & 73.87 & 84.84 & 29.51 \\
    \textit{\textbf{- GPT-4o}} & \textbf{54.10} & \textbf{72.44} & \textbf{76.03} & \textbf{86.89} & \textbf{32.28} \\
    \textit{- Mistral-Nemo} & 44.47 & 67.01 & 70.08 & 82.38 & 27.25 \\
    \bottomrule
    \multicolumn{6}{l}{
    \footnotesize
      \makecell[l]{
        \textbf{*}: EM scores are based on HiT@1 retrieval used to augment answer generation. \\
        Standard RAG and Table Retrieval do not use LLMs during retrieval. \\
        $\spadesuit$: These EM score reported in order: \textbf{GPT-4o-mini, GPT-4o, Mistral-Nemo}.
      }
    }
    \end{tabular}
    \label{tab:retrieval_baseline}
\end{table}

We execute QA tasks as an instance based on the most relevant document $ D^* $ provided by the Cross-Modal Retrieval Module. Although LLMs demonstrate strong performance in natural language understanding, performing QA on heterogeneous documents remains a challenging task. To more accurately extract relevant information from heterogeneous documents and leverage the information to answer the questions, we introduce a novel prompting strategy for QA tasks with \ourmodel{} called \textbf{RECAP} (\textbf{R}estate, \textbf{E}xtract, \textbf{C}ompute, \textbf{A}nswer, \textbf{P}resent), inspired by prompt strategies such as CoT~\cite{CoT} and EEDP~\cite{srivastava-etal-2024-evaluating} that involve step-by-step analytical reasoning, as well as PoT~\cite{PoT} which generates executable programs to solve complex mathematical operations. RECAP guides the LLM through a five-step linear process for reasoning and answering. Among these five steps, the \textbf{Compute} step plays a central role in improving answer quality, as it explicitly directs the model to reason and output intermediate logical steps or calculations when applicable.

Formally, as defined in Section~\ref{sec: formolize}, the \RecapModule{} is represented as $A = \text{Inference}(Q, D^*)$, where $ A $ is the final answer produced by the RECAP strategy. 
The five-step workflow of RECAP is as follows: 
\begin{enumerate}
    \item \textbf{Restate the question:} Clarify the question requirements and identify its type (e.g., numerical, fact-based, multi-hop). 
    \item \textbf{Extract relevant data:} Extract and list all content from $D^*$ that is directly relevant to the question.
    \item \textbf{Perform logical reasoning:} Guide the LLM to perform explicit reasoning or computations based on the extracted information. If applicable, the LLM is prompted to write out the full formula used in the computation. This step is crucial for improving accuracy in questions requiring logical operations or quantitative analysis. 
    \item \textbf{Generate final answer:} Generate a final answer based on the above step. If needed, an external calculator uses the formula provided to validate or refine the answer. A rule-based mechanism then selects the more appropriate of the two.
    \item \textbf{Present selected answer:} Return the selected answer in a clear and concise format.
\end{enumerate}


Unlike methods like Self-Consistency~\cite{self-consistency}, which improve accuracy by generating multiple reasoning paths, RECAP adopts a single-turn dialogue strategy, with its logical reasoning and formula provision originating from the same reasoning path, offering advantages in token efficiency for the LLM. 

\begin{table*}[htbp]
\centering
\setlength{\tabcolsep}{3.5pt} 
\caption{Comparison of RECAP in \ourmodel{} Document-QA Performance (EM Score\%) with Various Baselines}
  \begin{tabular}{l|c|c|c|c|c|c|c|c}
    \toprule
      & \multicolumn{8}{c}{\textbf{Models}} \\
    \cmidrule(lr){2-9}
    \textbf{Method} & \textbf{GPT-4o} & \textbf{GPT-4o mini} & \textbf{Gemini-2.0} & \textbf{Qwen-Plus} & \textbf{Qwen2.5-32B} & \textbf{Qwen2.5-7B} & \textbf{Mistral-Nemo} & \textbf{Llama3.1-7B} \\
    \midrule
    Direct         & 35.34 & 23.55 & 26.64 & 34.83 & 21.80 & 8.02  & 10.03 & 15.79 \\ 
    CoT            & 46.87 & 36.84 & 38.01 & 50.20 & 48.37 & 31.58 & 25.06 & 25.31 \\
    EEDP           & 47.37 & 37.59 & 40.37 & 51.23 & 48.62 & 28.57 & 27.57 & 20.80 \\
    PoT            & 61.90 & 55.64 & 51.84 & 60.45 & 55.89 & 37.84 & 34.09 & 29.32 \\
    \textbf{RECAP} & \textbf{64.66} & \textbf{58.15} & \textbf{60.25} & \textbf{64.45} & \textbf{64.16} & \textbf{45.36} & \textbf{44.11} & \textbf{40.60} \\ 
    \bottomrule
  \end{tabular}
  \label{tab:DocQA_baseline}
\end{table*}

\section{Experiment}
\label{Experiment}

\subsection{Experimental Setup}
To evaluate the effectiveness of \taskname{}, we compare \ourmodel{} against several representative baselines.
For the retrieval task, we include several representative baselines:
\begin{itemize}
    \item \textbf{Standard RAG}: A basic retrieval-augmented generation framework that retrieves documents using dense similarity search with embeddings from pre-trained models such as DPR, without table-specific optimization.
    \item \textbf{LangChain}~\cite{mavroudis2024langchain}: A semi-structured retrieval pipeline that processes tabular and textual data by summarizing content with LLMs and conducting similarity-based retrieval.
    \item \textbf{SelfRAG}~\cite{asai2023selfrag}: A self-reflective retrieval framework that dynamically retrieves passages and uses reflection tokens to iteratively assess and refine outputs based on the model’s own feedback.
    \item \textbf{RAG-Critic}~\cite{dong-etal-2025-criticrag}: A framework based on a hierarchical error system that aligns a specialized error-critic model to guide an agentic workflow, facilitating error-driven self-correction.
    \item \textbf{Table Retrieval}~\cite{chen-etal-2024-table}: A method tailored for table retrieval that discovers join relationships and applies mixed-integer programming-based re-ranking to optimize both table-query and table-table relevance.
\end{itemize}

For the QA task, we consider several widely used reasoning baselines demonstrated strong performance in answering structured and semi-structured questions:
\begin{itemize}
    \item \textbf{Direct}: A simple method where the model directly generates an answer without explicit reasoning or intermediate steps.
    \item \textbf{COT}~\cite{CoT}: A reasoning-based prompting strategy that guides the model to generate intermediate reasoning steps, improving its ability to solve complex problems.
    \item \textbf{EEDP}~\cite{srivastava-etal-2024-evaluating}: A multi-step reasoning framework that decomposes the answer process into explanation, refinement, and decision phases to enhance accuracy.
    \item \textbf{POT}~\cite{PoT}: A technique that prompts the LLM to produce a structured program for reasoning, which is then executed externally to yield precise results, especially for numerical problems.
\end{itemize}

We adopt HiT@K and Exact Match (EM) as evaluation metrics for retrieval performance and QA accuracy, respectively. 
To evaluate efficiency, we employ token usage and wall-clock time as our metrics.
Further details are available in Appendix~\ref{metrics}.
We employ GPT-4o throughout our pipeline to align with model used by baselines, including heterogeneous document representation, LLM-based retrieval, and RECAP execution.
For embedding-based retrieval, we use OpenAI’s text-embedding-3-large model as the encoder.
According to Appendix~\ref{Appendix:hyperparameter}, we set the parameters of \ourmodel{} to $n=40$ and $m=60$ for ensemble retrieval.
For the QA instance RECAP, the number of examples for all prompt strategies is uniformly set to 1.
We evaluate the RECAP instance using GPT-4o, GPT-4o mini, Gemini-2.0-Flash and three open-source large language models, Qwen2.5-32B-Instruct, Mistral-Nemo-Instruct-2407, Llama-3.1-8B-Instruct.

\subsection{Comparison with Baselines}
In this section, we present a comprehensive analysis of \ourmodel{} compared with several strong retrieval-augmented baselines, across the \ourdataset{} dataset to evaluate effectiveness and efficiency.

\subsubsection{Effectiveness Comparison}

\textbf{Retrieval Performance Comparison with Baselines.}
\zc{Table~\ref{tab:retrieval_baseline} presents the retrieval performance of \ourmodel{} compared to several baselines across three different LLM backbones.}
\zc{Across all HiT@K metrics (K=1, 3, 5, 10), \ourmodel{} consistently achieves the best results, reaching a HiT@1 score of 54.1\% on GPT-4o, significantly outperforming the strongest baseline (Table Retrieval) by 46\%. 
The fine-grained row-and-column-level summarization, combined with the two-stage retrieval strategy, achieves substantial improvements in retrieval accuracy for complex heterogeneous documents.}

\textbf{Document-QA Accuracy Comparison with Baselines.}
\zc{Using RECAP as the QA instance module for each retrieval method, we evaluate document-level QA accuracy based on EM scores.
According to the EM results in Table~\ref{tab:retrieval_baseline}, \ourmodel{} also achieves the highest QA accuracy, with an EM score of 32.28\% which corresponds to a 27.1\% improvement over the second-best baseline, Table Retrieval.}

\zc{In addition, we separately evaluate RECAP on the QA task to assess its standalone effectiveness compared to other table reasoning baselines on \ourdataset{}.
As shown in Table~\ref{tab:DocQA_baseline}, RECAP consistently outperforms all competing methods across five LLMs, demonstrating its strong capability in handling heterogeneous document QA.
Notably, by leveraging external calculation functions for arithmetic reasoning, RECAP improves the accuracy of computation-intensive questions while avoiding the instability often observed in program-executed strategies like PoT.
Furthermore, its decomposition-based prompting strategy enables multi-step reasoning, especially in scenarios with heterogeneous document.}

These results highlight the effectiveness of our retrieval and reasoning framework in improving answer accuracy for heterogeneous documents retrieval in downstream QA tasks.

\subsubsection{Efficiency Comparison}
We evaluate the efficiency of \ourmodel{} in terms of token consumption and wall-clock time using GPT-4o.
As detailed in Table~\ref{tab:cost_time_comparison}, \ourmodel{} strikes a balance between cost and performance. While it naturally incurs higher latency than single-step baselines (e.g., Standard RAG) due to its multi-hop reasoning mechanism, it is significantly more efficient than other iterative frameworks. Specifically, \ourmodel{} reduces generation tokens by approximately 41\% compared to Self-RAG and 59\% compared to RAG-Critic. 
Notably, \ourmodel{} achieves the lowest preprocessing token costs among all retrieval baselines, as well as lower generation token costs compared to iterative baselines, demonstrating its cost-efficiency.
Overall, \ourmodel{} achieves a superior trade-off between effectiveness and efficiency, delivering high reasoning performance while maintaining significantly low computational overhead.

\begin{table}[htbp]
  \centering
  \caption{Comparison of Token Cost and Time on DocRAGLib}
  \begin{tabular}{l|c|c|c}
    \toprule
    \textbf{Method} & \textbf{Pre. Tok.} & \textbf{Gen. Tok.} & \textbf{Time(s)} \\
    \midrule
    Standard RAG    & 0       & 4614.5  & 5.26  \\
    Table Retrieval & 1339.6  & 5371.9  & 3.81  \\
    LangChain       & 5212.5  & 6944.4  & 3.37  \\
    Self-RAG        & 2551.3  & 24165.9 & 15.47 \\
    RAG-Critic      & 4917.3  & 34989.8 & 23.89 \\
    \midrule
    \textbf{MixRAG} & \textbf{1108.3} & \textbf{14189.4} & \textbf{11.76} \\
    \bottomrule
  \end{tabular}
  \label{tab:cost_time_comparison}
\end{table}

\subsection{Ablation Study}

\begin{table*}[htbp]
    \centering
    \caption{Performance Evaluation of \ourmodel{} with Different Table Summary Level}
    \begin{tabular}{c|c|c|c|c|c}
    \toprule
    \textbf{Heterogeneous Document Representation} & \multicolumn{4}{c|}{\textbf{HiT@K(\%)}} & \multirow{2}{*}{\textbf{EM*(\%)}} \\
    \cmidrule(lr){2-5}
    \textbf{Table Summary Level} & \textbf{K=1} & \textbf{K=3} & \textbf{K=5} & \textbf{K=10} \\
    \midrule
    Table Level Summary   & 36.27 & 48.05 & 59.63 & 74.28 & 23.83 \\
    General RCL Summary   & 39.84 & 55.02 & 68.03 & 79.81 & 26.25 \\
    H-RCL Summary         & \textbf{54.10} & \textbf{72.44} & \textbf{76.03} & \textbf{86.89} & \textbf{32.28} \\
    \bottomrule
    \multicolumn{6}{l}{\footnotesize *Note: EM scores are based on HiT@1 retrieval using RECAP.}
    \end{tabular}
    \label{tab:our_method_evaluation}
\end{table*}
To systematically investigate the effectiveness of each key component in \ourmodel{}, we conduct comprehensive ablation studies based on the DocRAGLib dataset. 
The studies include variations in the heterogeneous document representation strategies and \ourmodel{} modules.

\textbf{Impact of Heterogeneous Document Representation Strategy in \ourmodel{}.}
To evaluate the impact of different table summarization methods in heterogeneous document representation module, we compare three variants: Table Level Summary, General RCL Summary, and H-RCL Summary. 
As shown in Table~\ref{tab:our_method_evaluation}, the H-RCL Table Summary strategy achieves the best performance among all heterogeneous document representation strategies.
The General RCL Summary ranks second, showing clear gains over the traditional Table Level Summary by providing more explicit representations of rows and columns.
H-RCL improves HiT@1 by 47\% and EM by 28\% compared to the Table Level Summary, demonstrating the importance of capturing multi-level table structures.

\textbf{The Table Level Summary} performs poorly because it tends to over-compress complex tabular data into flat textual descriptions. It sacrifices fine-grained details and row-column relationships, making it difficult for the retriever to locate precise evidence.
\textbf{The General RCL} strategy ranks second. It provides a clearer representation of rows and columns, reducing ambiguity and improving retrieval accuracy, which in turn enhances QA performance by enabling more precise document matching. 
However, it treats tables as flat grids and struggles to capture the nested dependencies found in hierarchical tables.
Building on this, \textbf{the H-RCL strategy} addresses these limitations by further capturing multi-level table structures and incorporating hierarchical path information. This design preserves the full lineage of cell values, significantly boosting both retrieval accuracy and the model’s ability to perform complex reasoning over structured data.
Overall, these results demonstrate the effectiveness of the H-RCL design in improving both retrieval and reasoning capabilities for heterogeneous documents.


\begin{table}[htbp]
\small
  \centering
  \caption{Ablation Study Result of \ourmodel{}}
  \begin{tabular}{l@{\hspace{0.8em}}c@{\hspace{0.8em}}c}
    \toprule
     & \textbf{HiT@1(\%)} & \textbf{EM(\%)} \\
    \midrule
    \textbf{Ours (\ourmodel{})}      & \textbf{54.10} & \textbf{32.38} \\
    \textbf{- w/o Ensemble Retrieval}&                 &                \\
    \hspace{1em}only Embedding       & 47.34          & 27.77         \\
    \hspace{1em}only BM25            & 35.45          & 21.00         \\
    \textbf{- w/o Two-stage Retrieval*}&                &                \\
    \hspace{1em}only Embedding       & 27.87          & 21.31         \\
    \hspace{1em}only BM25            & 15.57          & 13.11         \\
    \textbf{- w/o \textit{R} of \textit{RECAP}}     & 54.10          & 31.45         \\
    \textbf{- w/o \textit{E} of \textit{RECAP}}     & 54.10          & 30.63         \\
    \textbf{- w/o \textit{C} of \textit{RECAP}}     & 54.10          & 25.61         \\
    \bottomrule
    \multicolumn{3}{l}{\footnotesize *Note: w/o both Ensemble Retrieval and LLM-based Retrieval.}
  \end{tabular}
  \label{tab:ablation_study}
\end{table}

\textbf{Impact of Retrieval and QA Module.}
To investigate the effectiveness of each key component in \ourmodel{}, we conduct the ablation study using \ourdataset{} dataset.
Table~\ref{tab:ablation_study} shows the results of the Retrieval and QA module ablation study. 
Replacing Ensemble Retrieval with either Embedding or BM25 significantly reduces HiT@1, with Embedding achieving better results due to its stronger semantic alignment capabilities.
Further removing LLM-based retrieval leads to an additional decline of 19\% and 20\% in HiT@1 for Embedding and BM25, respectively, confirming the advantage of the two-stage retrieval framework over single-stage methods.
Moreover, disabling external calculation in RECAP results in a noticeable drop in EM, underscoring the importance of precise formula-based reasoning for improving answer accuracy.
Overall, these results validate the effectiveness of each component within \ourmodel{} in enhancing both retrieval precision and answer accuracy.


\subsection{Scalability Analysis}
\begin{figure}[htbp]
  \centering
  \includegraphics[width=0.9\linewidth]{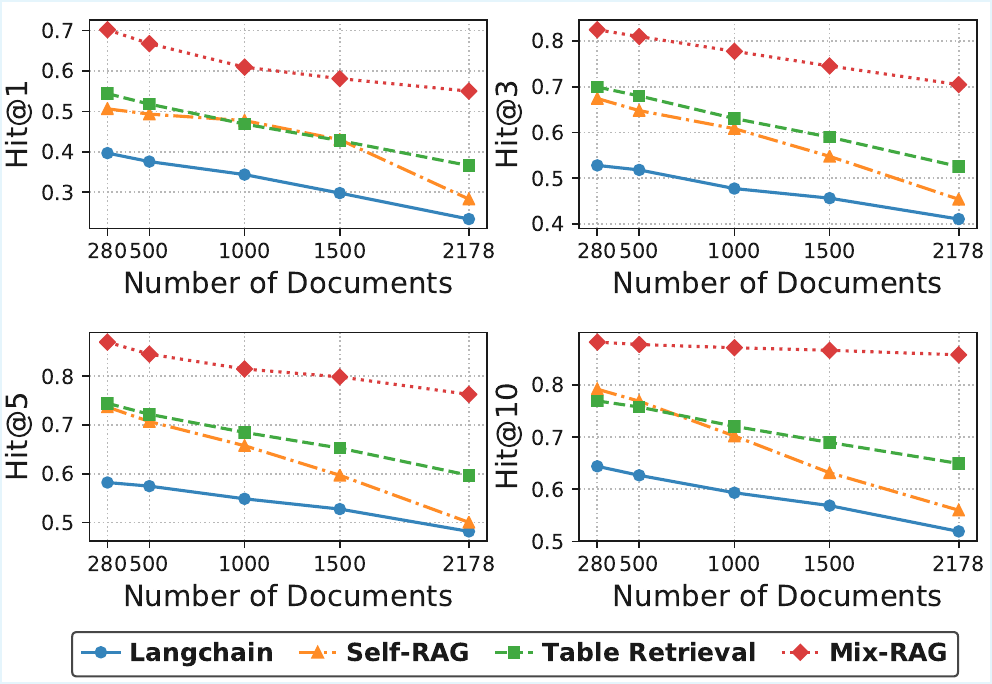}
  \caption{Scalability Analysis of Retrieval Performance across Different Corpus Sizes.}
  \label{fig:scalability}
\end{figure}

\zc{We evaluate the scalability of \ourmodel{} by comparing its retrieval performance with several baselines across different document corpus sizes in GPT-4o.
Standard RAG is excluded from this analysis due to its consistently low retrieval accuracy.}
As shown in Figure~\ref{fig:scalability}, \ourmodel{} consistently maintains the highest HiT@K scores (K=1, 3, 5, 10) across all corpus scales, demonstrating its ability to preserve retrieval precision even as the corpus size increases.
While baseline methods experience significant drops in performance as the document corpus increases, \ourmodel{} exhibits the slowest rate of degradation, particularly in HiT@10, where its performance remains notably stable even with a larger document corpus.
The retrieval stability reflects the scalability of \ourmodel{} in handling growing document complexity.
These results confirm that \ourmodel{} effectively maintains retrieval precision even with larger and more complex document sets, showcasing its superior scalability compared to existing methods.

\subsection{Failure Case Study}

In this section, we analyze the failure cases in our experiments.
We categorize these failure cases into three main types.

\zc{\textbf{Retrieval Failure Due to Incomplete Question.} 
When questions are unclear or lack explicit references to key information within the documents, \ourmodel{} struggles to identify content that is strongly related to the question, leading to retrieval wrong document.
This problem is particularly common in RAG systems, as they heavily rely on explicit question-document relationships.
Without clear references in the question to guide the retrieval, it may retrieval irrelevant document, which in turn impacts the accuracy of the subsequent task.
}

\zc{\textbf{Reasoning Failure Due to Complex Multi-Step Tasks.} 
When the question requires complex multi-step reasoning beyond the LLM's capabilities in a single prompt, the LLM may fail to generate the correct answer.
It is common in tasks involving numerical calculations and logical deductions.
}

\zc{\textbf{Reasoning Failure Due to Ambiguity in Reasoning Paths.}
When questions have multiple valid interpretations, our \RecapModule{} may choose an incorrect reasoning path, resulting in errors in the final answer.
This issue is common in LLM-based methods, particularly when tasks have multiple possible answers or reasoning approaches.}

\zc{In our future work, we plan to design a question decomposition approach to break down the complex questions into sub-questions, improving retrieval accuracy by ensuring the clarity of each sub-question.
Additionally, the decomposition will explicitly display the reasoning paths for LLM execution, supporting multi-step reasoning through multiple calls, which enhances the LLM's ability to handle complex tasks. 
Furthermore, we aim to introduce validation for reasoning paths to reduce ambiguity and guarantee the selection of the most appropriate path, ultimately improving the accuracy and reliability of the reasoning process.
By addressing these failure cases and incorporating these improvements, we aim to improve the performance and robustness of \ourmodel{} in future work.}
\section{Conclusion}
\label{Conclusion}
In this paper, we investigate the challenge of question answering over heterogeneous documents. 
To support this task, we introduce \ourdataset{}, a benchmark dataset containing over 2,000 heterogeneous documents and over 4,000 QA pairs. 
To effectively address the challenges in this task, we propose the \ourmodel{} framework. 
MixRAG integrates hierarchical table summarization, cross-modal retrieval with LLM-based reranking, and RECAP-guided multi-step reasoning to support RAG over heterogeneous documents.
Extensive experiments show that \ourmodel{} outperforms existing baselines in retrieval precision and QA accuracy, establishing new state-of-the-art results for heterogeneous document grounding while maintaining a superior trade-off between effectiveness and efficiency.

\begin{acks}
We would like to thank Haoyu Dong for his valuable assistance and constructive suggestions, which greatly improved this work.
\end{acks}



\bibliographystyle{ACM-Reference-Format}
\bibliography{sample-base}

\appendix

\section{Dataset}
\label{Dataset_Appendix}


\subsection{Data Cleaning}
The raw data consists of heterogeneous documents and QA pairs. This section outlines the data cleaning process, focusing on the strategies to refine both documents and QA pairs.

For document refinement, we reduce noise in table contexts extracted from web pages by addressing issues such as irregular delimiters and unrelated sentences. Noise removal is performed through manual cleaning, which involves (1) reformatting inconsistent delimiters and (2) removing irrelevant text segments.

\begin{figure}[htbp]
    \centering
    \includegraphics[width=1\linewidth]{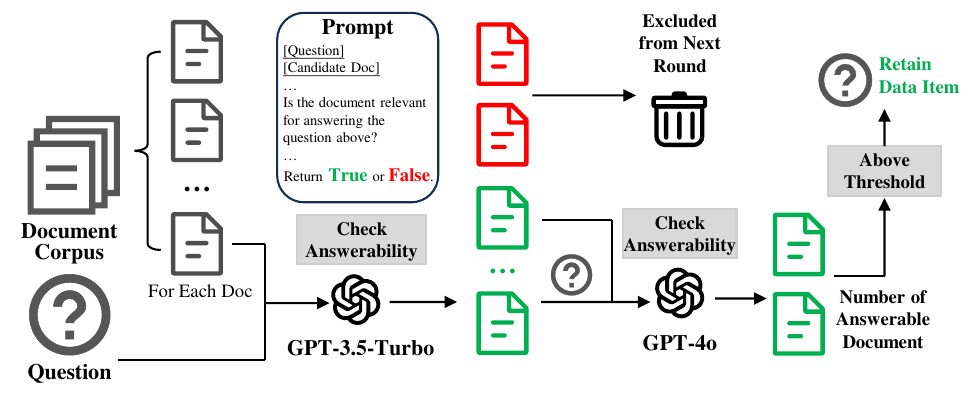}
    \caption{Pipeline for ambiguity resolution during QA pair construction}
    \label{fig:data_clean}
\end{figure}

To ensure clarity in the QA pairs, we further resolve ambiguities arising when multiple documents could potentially answer the same question. The process begins with rule-based filtering to identify candidate documents. Each candidate is then evaluated for relevance by pairing it with the question and inputting the pair into GPT-3.5 Turbo and GPT-4o. Relevance judgments are tracked iteratively, and if the number of remaining documents falls below a predefined threshold, the question is considered unambiguous and the corresponding QA pair is retained. As shown in Figure~\ref{fig:data_clean}, this process systematically eliminates ambiguous questions and ensures the overall quality of the dataset.

\cqy{In addition to the automated cleaning pipeline described above, we conducted a manual evaluation stage to further enhance data quality. Evaluation team members reviewed a random sample of the dataset; each member examined at most 100 heterogeneous documents together with all of their associated QA pairs. This manual audit led to the removal of 249 documents and 370 QA pairs. The principal reasons for filtering were: (1) ambiguous question formulations (in these cases only the question itself was discarded); (2) mismatches between textual passages and table information; (3) missing or incomplete textual content; and (4) irregular table structures (e.g., merged cells that disrupt the hierarchical layout). For the latter three issues, because the documents contained irreparable defects, we deleted the entire document along with all related questions. This additional manual quality-control step markedly improved the dataset’s overall accuracy, consistency, and readability.}

\subsection{Data Statistics}

After the data collection and cleaning process, 2,178 heterogeneous documents and 4,468 QA pairs are obtained. To enable further exploration of pre-trained model-based methods for the \taskname{} task, the 4,468 QA pairs are split into three subsets: train (2,990 QA pairs), dev (502 QA pairs), and test (976 QA pairs). To ensure consistency, the 2,178 heterogeneous documents are merged into a single collection, without further division.

As shown in Table~\ref{tab:statistics}, statistical analysis of the \ourdataset{} dataset reveals that: (1) The average textual length of each document is 1,453.05 words, and the average number of tables is 3.85, with 88.71\% of documents containing at least two tables. (2) The QA pairs are categorized by evidence source: 9.02\% of questions rely solely on text, 50.78\% on tables, and 40.20\% on both.
Additionally, we analyze the numerical scale and precision of answers in the \ourdataset{} dataset. Specifically, 19.36\% of answers contain values exceeding 10,000, 4.05\% exceed 100,000, and 34.29\% are decimals with more than three significant digits. These statistics highlight the complexity and diversity of the dataset, offering a challenging benchmark for future research.
\begin{table}[t]
    \centering
    \caption{Statistical Analysis of the \ourdataset{} Dataset}
    \begin{tabular}{ll}
            \toprule
            \textbf{Indicator}                                & \textbf{Value}      \\ 
            \midrule
            \textbf{Total Document Number}              & \textbf{2,178}              \\
            \quad \quad Avg. Word Number per Document                       & 1,453.05            \\
            \quad \quad Avg. Table Number per Document        & 3.85                \\
            \quad \quad Documents with >2 Tables       & 1,932 (88.71\%)             \\ 
            \midrule
            \textbf{Total QA Pairs Number}                                    & \textbf{4,468}               \\
            \quad \quad Train/Dev/Test                                       & 2,990/502/976               \\

            \textbf{Classification by Evidence Source}       &                     \\
            \quad \quad Text Only                                   & 403 (9.02\%)        \\
            \quad \quad Table Only                                  & 2,269 (50.78\%)     \\
            \quad \quad Both Text and Table                         & 1,796 (40.20\%)     \\ 
            \textbf{Answer Characteristics}                  &                     \\
            \quad \quad Decimal Places $\geq$3                      & 1,532 (34.29\%)             \\ 
            \quad \quad Numerical Values >10,000                    & 865 (19.36\%)             \\
            \quad \quad Numerical Values >100,000                   & 181 (4.05\%)              \\
        \bottomrule
    \end{tabular}
    \label{tab:statistics}
\end{table}

\section{Evaluation Metrics}
\label{metrics}

We adopt two standard metrics to evaluate the performance of the retrieval and QA modules:

\begin{itemize}
    \item \textbf{HiT@K}: Measures the proportion of questions for which the correct document is retrieved within the top-K candidates. This metric reflects the effectiveness of the retrieval module. Formally,
    \begin{equation}
        \text{HiT@K} = \frac{1}{N} \sum_{i=1}^{N} \mathbb{I} \left[ d_i^\ast \in \text{TopK}(q_i) \right],
    \end{equation}
    where $N$ is the total number of questions, $q_i$ is the $i$-th question, $d_i^\ast$ is its ground-truth document, and $\mathbb{I}[\cdot]$ is the indicator function.
    
    \item \textbf{Exact Match (EM)}: Calculates the percentage of predicted answers that exactly match the ground-truth answers. This metric evaluates the accuracy of the QA module. Formally,
    \begin{equation}
        \text{EM} = \frac{1}{N} \sum_{i=1}^{N} \mathbb{I} \left[ a_i = \hat{a}_i \right],
    \end{equation}
    where $a_i$ and $\hat{a}_i$ denote the ground-truth and predicted answers for question $q_i$, respectively.

    \item \textbf{Token Usage}: We measure the computational cost via the number of tokens consumed by the LLM. This metric is decomposed into \textit{preprocessing tokens} and \textit{generation tokens}.

    \item \textbf{Time}: We record the average end-to-end wall-clock time (in seconds) required to process a query.
\end{itemize}

\section{Hyperparameter Experiment}
\label{Appendix:hyperparameter}
In this section, we investigate the effects of two key hyperparameters within the Ensemble Retrieval module of \ourmodel{}. These include the number of documents $n$ selected by the BM25 retriever and the number of top documents $m$ selected by the embedding-based retriever.

The final number of chunks $k$, passed into the LLM-based retrieval stage, is constrained by the total number of candidates retrieved, i.e., $k \leq n+m$.
In practice, $k$ determines the number of document chunks provided as input to the language model, and thus directly affects both context length and inference quality. Considering the token limitations and reasoning capacity of LLMs, we empirically find that setting $k \approx 100$ strikes a balance between informativeness and model performance.

Based on this constraint, we conduct a grid search over different combinations of $n$ and $m$ such that $n+m=100$, where $k$ denotes the number of unique document chunks after deduplication. We then evaluate the impact of these configurations on retrieval performance.

\begin{table}[htbp]
\centering
\caption{Retrieval performance under different $(n, m)$ configurations with deduplicated document count $k$.}
\begin{tabular}{c|c|c|c}
    \toprule
    \textbf{BM25 ($n$)} & \textbf{Embedding ($m$)} & \textbf{Average of $k$} & \textbf{HiT Rate} \\
    \midrule
    70 & 30 & 96.73 & 0.9016 \\
    60 & 40 & 94.26 & 0.9344 \\
    50 & 50 & 93.18 & 0.9703 \\
    \textbf{40} & \textbf{60} & \textbf{93.62} & \textbf{0.9805} \\
    30 & 70 & 94.92 & 0.9487 \\
    \bottomrule
    \multicolumn{4}{l}{\footnotesize *HiT Rate: proportion of ground-truths in the $k$ deduplicated documents.}
\end{tabular}
\label{tab:hyperparam}
\end{table}

\begin{table}[htbp]
\small
\centering
\caption{Evaluation of the Reranking Performance of Different Models}
\begin{tabular}{lcccc}
\toprule
\textbf{Reranker} & \textbf{Model} & \textbf{HiT@1} & \textbf{HiT@3} & \textbf{HiT@5} \\
\midrule
LLM & GPT-4o & \textbf{0.73} & \textbf{0.79} & \textbf{0.86} \\
     & GPT-4o mini & \underline{0.63} & \underline{0.70} & 0.77 \\
     & GPT-3.5 Turbo & 0.51 & 0.57 & 0.69 \\
\midrule
Cross-Encoder & MiniLM-L-12-v2 & 0.56 & 0.67 & \underline{0.79} \\
              & MiniLM-L-6-v2  & 0.48 & 0.58 & 0.65 \\
              & MiniLM-L-2-v2  & 0.52 & 0.57 & 0.62 \\
\midrule
BGE-Reranker & reranker-v2-m3 & 0.58 & 0.66 & 0.72 \\
             & base-en-v1.5   & 0.53 & 0.55 & 0.56 \\
             & large-en-v1.5  & 0.49 & 0.54 & 0.56 \\
             & base-en        & 0.37 & 0.42 & 0.47 \\
\bottomrule
\end{tabular}
\label{tab:rerank_evaluation}
\end{table}

HiT Rate measures the proportion of cases where the ground-truth document appears within the $k$ deduplicated document chunks returned by Ensemble Retrieval, thereby reflecting the overall retrieval precision.
According to Table~\ref{tab:hyperparam}, the configuration with $n=40$ and $m=60$ achieves the highest retrieval performance, reaching a HiT Rate of 0.9805 with an average deduplicated document count of 93.62. 
This setting offers an effective balance between sparse and dense retrieval, leading to more comprehensive and relevant document coverage. 
We adopt $n=40$ and $m=60$ as the default hyperparameter setting for Ensemble Retrieval in our subsequent experiments.

\section{Evaluation of the Reranking Performance of Different Models}
\label{Appendix:Evaluation_rerank}

In the LLM-based Retrieval stage, the input to the LLM is the Top k documents obtained after Ensemble Retrieval. At this stage, the LLM can be viewed as a reranking model. After reconstructing the Top k documents, it reorders them based on the similarity between the document content and the question, ultimately selecting the most relevant documents. In this section, we further evaluate the performance differences between the LLM and conventional reranking models in this fine-grained retrieval reranking task.

In order to benchmark the efficacy of our LLM-based reranking against established retrieval paradigms, we evaluate three distinct families of models: (1) Large Language Model (LLM) rerankers, (2) Cross-encoder rerankers, and (3) Bi-encoder embedding rerankers. For the LLM family, we select GPT-4o and its compact variant GPT-4o mini, as well as GPT-3.5 Turbo, due to their demonstrated ability to perform nuanced contextual matching and logical inference over long documents. These models ingest the entire reconstructed top-k candidate set and exploit autoregressive decoding to implicitly compare semantic relevance across documents.

In contrast, the cross-encoder models, including ms-marco-MiniLM series (e.g., MiniLM-L12-v2, MiniLM-L6-v2, and MiniLM-L2-v2), perform fine-grained token-level interactions via full‐attention between query and document, yielding highly calibrated relevance scores at the expense of greater computational cost. Finally, we include the BGE reranker series (e.g., bge-reranker-v2-m3, base-en-v1.5, large-en-v1.5, and base-en) to represent bi-encoder architectures that first encode query and document separately into fixed embeddings and then score relevance via cosine similarity. Together, these three classes span the methodological spectrum from generative inference to cross-attention scoring and pure embedding matching.

The document and question set used for this performance evaluation is a subset of \ourdataset, consisting of 100 questions and 280 documents, ensuring that each question has a corresponding document. Using the Ensemble Retrieval Module, we retrieve the corresponding Top k (k=100) documents for each question from the 280 documents. At this point, the HiT@100 for all questions is 0.98, meaning that for 98 of the questions, the Top 100 documents contain the unique document corresponding to each question.

In Table~\ref{tab:rerank_evaluation}, we present the evaluation results of different models in the fine-grained retrieval reranking task. The metrics shown in the table include HiT@1, HiT@3, and HiT@5, which measure the models' ability to correctly rank and further retrieve the top documents from the Top k documents. From the results, it is evident that GPT-4o performs the best across all evaluation metrics, particularly in scenarios requiring high ranking accuracy. Specifically, GPT-4o achieves a HiT@1 score of 0.73, surpassing GPT-4o mini by 15.9\% and the best-performing rerank model, ms-marco-MiniLM-L12-v2, by 25.9\%.

\end{document}